\documentclass[useAMS,usenatbib]{mn2e}
\usepackage{natbib}
\usepackage{epsfig}
\usepackage{psfig}
\usepackage{aastex_hack}
\usepackage{longtable}

\def\gsim { \lower .75ex \hbox{$\sim$} \llap{\raise .27ex \hbox{$>$}} }
\def\lsim { \lower .75ex \hbox{$\sim$} \llap{\raise .27ex \hbox{$<$}} }

\title[$\Lambda$CDM halo structure III: Universality and Asymptotic Slopes]
{The Inner Structure of $\Lambda$CDM Halos III: Universality and Asymptotic Slopes}
\author[J.~F.~Navarro et al.]
{J.~F.~Navarro$^{1,6}$,E.~Hayashi$^{1}$, C.~Power$^2$, A.~R.~Jenkins$^2$,
C.~S.~Frenk$^2$, \newauthor S.~D.~M.~White$^3$, V.~Springel$^3$, J.~Stadel$^4$
and T.~R.~Quinn$^5$\\ 
$^1$Department of Physics and Astronomy, University of
Victoria, Victoria, BC, V8P 1A1, Canada\\ 
$^2$ Institute for Computational
Cosmology, Department of Physics, University of Durham, South Road, Durham, DH1
3LE\\ 
$^3$ Max-Planck Institute for Astrophysics, Garching, Munich, D-85740,
Germany\\ 
$^4$ Institute for Theoretical Physics, University of Zurich, Zurich
CH-8057, Switzerland\\ 
$^5$ Department of Astronomy, University of Washington,
Seattle, WA 98195, USA\\ 
$^6$ Fellow of the Canadian Institute for Advanced
Research and of the J.S.Guggenheim Memorial Foundation}

\begin{document}










\maketitle

\begin{abstract}
We investigate the mass profile of $\Lambda$CDM halos using a suite of
numerical simulations spanning five decades in halo mass, from dwarf
galaxies to rich galaxy clusters. These halos typically have a few
million particles within the virial radius ($r_{200}$), allowing
robust mass profile estimates down to radii below $1\%$ of
$r_{200}$. Our analysis confirms the proposal of Navarro, Frenk \&
White (NFW) that the shape of $\Lambda$CDM halo mass profiles differs
strongly from a power law and depends little on mass.  The logarithmic
slope of the spherically-averaged density profile, as measured by
$\beta=-d\ln \rho/d\ln r$, decreases monotonically towards the center
and becomes shallower than isothermal ($\beta < 2$) inside a
characteristic radius, $r_{-2}$. The fitting formula proposed by NFW
provides a reasonably good approximation to the density and circular
velocity profiles of individual halos; circular velocities typically
deviate from best NFW fits by less than $10\%$ over the radial range
which is well resolved numerically.  On the other hand, systematic
deviations from the best NFW fits are also noticeable. Inside
$r_{-2}$, the profile of simulated halos becomes shallower with radius
more gradually than predicted and, as a result, NFW fits tend to
underestimate the dark matter density in these regions. This
discrepancy has been interpreted as indicating a steeply divergent
cusp with asymptotic inner slope, $\beta_{0}\equiv\beta(r=0)\sim
1.5$. Our results suggest a different interpretation. We use the
density and enclosed mass at our innermost resolved radii to place
strong constraints on $\beta_{0}$: density cusps as steep as
$r^{-1.5}$ are inconsistent with most of our simulations, although
$\beta_{0}=1$ is still consistent with our data. Our density profiles
show no sign of converging to a well-defined asymptotic inner power
law. We propose a simple formula that reproduces the radial dependence
of the slope better than the NFW profile, and so may minimize errors
when extrapolating our results inward to radii not yet reliably probed
by numerical simulations.
\end{abstract}



\section{Introduction}
\label{sec:intro}

Disk galaxy rotation curves; strong gravitational lensing by galaxies
and clusters; the dynamics of stars in elliptical galaxies and of gas
and galaxies in clusters; these are just examples of the various
luminous tracers that probe the inner structure of dark matter
halos. Such observations place strong constraints on the distribution
of dark matter in these highly non-linear regions that may be
contrasted directly with theoretical predictions for halo
structure. Given the sensitivity of such predictions to the nature of
the dark matter, these observational constraints constitute provide
tests that may question or even rule out particular models of dark
matter.

Robust prediction of the inner structure of cold dark matter (CDM)
halos is a quintessential N-body problem, albeit one of considerable
complexity due to the large overdensities and, consequently, the short
crossing times involved. Indeed, only recently have computational
capabilities improved to the point of allowing realistic simulation of
the regions which house the luminous components of individual
galaxies.

This work builds upon the pioneering efforts of Frenk et al (1988), Dubinski and
Carlberg (1991), and Crone et al (1994), among others, which led to the
identification of a number of key features of the structure of dark matter halos
assembled by hierarchical clustering. One important result of this early work
concerns the absence of a well defined central ``core'' of constant density in
virialized CDM halos.  In this sense, dark matter halos are ``cuspy'': the dark
matter density increases apparently without bounds toward the center of the halo.

A second result concerns the remarkable similarity (``universality'')
in the structure of dark matter halos of widely different mass. This
was first proposed by Navarro, Frenk \& White (1996, 1997; hereafter
NFW), who suggested a simple fitting formula to describe the
spherically-averaged density profiles of dark matter halos,
\begin{equation}
\rho(r)={\rho_s \over (r/r_s) (1+(r/r_s))^{2}},
\label{eq:nfw}
\end{equation}
where $\rho_s$ and $r_s$ are a characteristic density and radius,
respectively. The larger the mass of a halo, the lower its characteristic
density, reflecting the lower density of the universe at the (later) assembly
time of more massive systems.

Further simulation work of similar numerical resolution (see, e.g.,
Cole \& Lacey 1996, Huss et al 1999) provided support for the NFW
conclusions, although small but systematic differences began to emerge
as the numerical resolution of the simulations improved (see, e.g.,
Moore et al 1999, hereafter M99, Ghigna et al 2000, Fukushige \&
Makino 1997, 2001, 2003). These authors reported deviations from
eq.~\ref{eq:nfw} that increase systematically inward, and thus are
particularly noticeable in high-resolution simulations.  In
particular, Fukushige \& Makino (2001) reported that NFW fits tend to
underestimate the dark matter density within the scale radius
$r_s$. M99 reached a similar conclusion and interpreted this result as
indicating a density cusp steeper than that of the NFW profile. These
authors preferred a modified fitting function which diverges as
$r^{-1.5}$ near the center,
\begin{equation}
\rho(r)={\rho_M \over (r/r_M)^{1.5} (1+(r/r_M)^{1.5})}.
\label{eq:mooreetal}
\end{equation}
One should note, however, that there is no consensus amongst N-body
practitioners for such modified profile (see, e.g., Klypin et al 2001
and Power et al 2003, hereafter P03), and that there is some work in
the literature suggesting that the central density cusp might actually
be {\it shallower} than $r^{-1}$ (Subramanian et al 2000; Taylor \&
Navarro 2001; Ricotti 2003).

This unsettled state of affairs illustrates the difficulties
associated with simulating the innermost structure of CDM halos in a
reliable and reproducible manner. The high density of dark matter in
such regions demands large numbers of particles and fine time
resolution, pushing to the limit even the largest supercomputers
available at present. As a result, many of the studies mentioned above
are either of inadequate resolution to be conclusive or are based on
results from a handful of simulations where computational cost
precludes a systematic assessment of numerical convergence.

Establishing the detailed properties of the central density cusp, as
well as deriving the value of its asymptotic central slope, are
important for a number of reasons. For example, steeper cusps place
larger amounts of dark matter at the center, exacerbating the
disagreement with observations that suggest the presence of a constant
density core in low surface brightness galaxies or in strongly barred
galaxies (Moore 1994, Flores \& Primack 1994, McGaugh \& de Blok 1998,
Debattista and Sellwood 1998, van den Bosch et al 2000). Steep cusps
would also be important for direct detection experiments for dark
matter, as a possible gamma-ray annihilation signal of WIMPS at the
Galactic center would be particularly strong for $r^{-1.5}$ cusps
(Calc\`aneo-Rold\'an \& Moore 2000; Taylor \& Silk 2003; Stoehr et al
2002).

Finally, the detailed structure of the central cusp is not the only
focus of contrasting claims in the literature. For example, the
``universality'' of CDM halo structure has been questioned by
Jing \& Suto (2000), who find that the slope of the density profile at
a fixed fraction of the virial radius steepens towards lower halo
masses. Klypin et al (2001), on the other hand, point out that such a
systematic trend is entirely consistent with universality as
originally claimed by NFW, and just reflects the mass dependence of
halo characteristic density.

We address these conflicting issues here using a suite of nineteen
high-resolution simulations of the formation of halos in the standard
$\Lambda$CDM cosmogony. Halo masses are chosen in three main groups:
``dwarf'' halos with $M_{200} \sim  10^{10} \, h^{-1} \,
\,M_{\odot}$, ``galaxy'' halos with $M_{200}\sim 10^{12} \, h^{-1}
\,M_{\odot}$ and ``cluster'' halos of mass $M_{200}\sim 10^{15} \,
h^{-1} \, M_{\odot}$. This allows us to gain insight into the effects
of cosmic variance at each mass scale, as well as to explore the mass
dependence of the structure of $\Lambda$CDM halos. We define the mass
of a halo to be that contained within its virial radius, that is,
within a sphere of mean density contrast $200${\footnote{We use the
term `density contrast' to denote densities expressed in units of the
critical density for closure, $\rho_{\rm crit}=3H^2/8\pi G$. We
express the present value of Hubble's constant as $H(z=0)=H_0=100\, h$
km s$^{-1}$ Mpc$^{-1}$}}.

This paper is organized as follows. Section~\ref{sec:numexp} describes
briefly the numerical simulations; \S~\ref{sec:res} discusses our main
results; and we summarize our conclusions in \S~\ref{sec:conc}.

\section{Numerical Experiments}
\label{sec:numexp}

The numerical set up of our simulations follows closely the procedure
described by P03, where the interested reader may find a thorough
discussion of our initial conditions generating scheme, the choice of
N-body codes and integrators, as well as the criteria adopted to
optimize the choice of the numerical parameters of the
simulations. For completeness, we include here a brief discussion of
the main numerical issues, but refer the reader to P03 for a more
detailed discussion.

\subsection{N-body codes}

The simulations reported in this paper have been performed using two
parallel N-body codes: {\tt GADGET}, written by Volker Springel
(Springel, Yoshida \& White 2001), and {\tt PKDGRAV}, written by
Joachim Stadel and Thomas Quinn (Stadel 2001). As discussed in P03,
both codes give approximately the same results for appropriate choices
of numerical parameters, and neither code seems obviously to
outperform the other when similar numerical convergence criteria are
met.

\subsection{Cosmological Model}
\label{ssec:cosmod}

We adopt a flat, $\Omega_0=0.3$ $\Lambda$CDM cosmological model whose
dynamics is dominated at present by a cosmological constant,
$\Omega_{\Lambda}=0.7$.  The matter power spectrum is normalized so
that the present linear rms amplitude of mass fluctuations in spheres
of radius $8\, h^{-1}$ Mpc is $\sigma_8=0.9$. We assume a linear
fluctuation power spectrum given by the product of the square of the
appropriate CDM transfer function, $T^2(k)$, and a Harrison-Zel'dovich
primordial power spectrum (i.e. $P(k) \propto k$).

\subsection{Parent Simulations}
\label{ssec:parentsim}

The halo samples were drawn from three different $\Lambda$CDM
cosmological ``parent'' simulations. Table~\ref{tab:numexp} lists the
main numerical parameters of each of these simulations: $L_{\rm box}$
is the size of the cosmological box, $z_i$ is the initial redshift,
$m_p$ is the particle mass, and $\epsilon$ is the softening parameter,
assumed fixed in comoving coordinates.

The dwarf, cluster, and most of the galaxy halos, were extracted from
the simulations $\Lambda$CDM-512 (Yoshida, Sheth \& Diaferio 2001) and
SGIF-128.  These two parent simulations, both carried out within the
Virgo Consortium, used the CDM transfer function given by CMBFAST
(Seljak \& Zaldarriaga 1996), assuming $h=0.7$ and $\Omega_b=0.04$.
This transfer function is well fit by the Bardeen et al (1986) fitting
formula with a value of $0.17$ for the shape parameter $\Gamma$. Three
of the galaxy halos (G1-G3, see Table~\ref{tab:halpar}) were extracted
from the parent simulation described by Eke, Navarro \& Steinmetz
(2001, labeled ENS01 in Table~\ref{tab:numexp}). That simulation used the
Bardeen et al (1986) fitting formula for the CDM transfer function,
with $h=0.65$ and $\Gamma=0.2$.

\subsection{Initial Conditions}
\label{ssec:ics}

Since completing the numerical convergence tests reported in P03, we
have developed a more flexible and powerful set of codes for setting
up the resimulation initial conditions. This resimulation software
enables us to iterate the procedure in order to ``resimulate a
resimulation'', an important step for setting up appropriate initial
conditions for dwarf halos.  The basic methodology employed is very
similar to the methods described in P03, with just a few minor
differences.  Galaxy halos G1-G3 were selected from the ENS01
simulation and their initial conditions were created using the
software described in P03. All of the other halos were set up with the
new codes, following the procedure we describe below.

The first stage is to carry out, up to the redshift of interest
(typically $z=0$), a ``parent'' simulation of a large, representative
volume of a $\Lambda$CDM universe. These parent simulations are used
to select halos targeted for resimulation at higher resolution.  Once
a halo has been selected for resimulation at $z=0$, we trace all
particles within a sphere of radius $\sim 3\, r_{200}$ to the
$z=\infty$ ``unperturbed'' configuration. We then create a set of
initial conditions with much higher mass resolution in the volume
occupied by the halo particles, and resample the remainder of the
periodic box at lower resolution, taking care to retain sufficient
resolution in the regions surrounding the halo of interest so that
external tidal forces acting on the high-resolution region are
adequately represented.

The procedure involves two main steps. Firstly, we set up a uniform
multi-mass distribution of particles to approximate the particle
positions in the high-resolution region at $z=\infty$. This is
accomplished by arranging particles either in a cubic grid or as a
``glass'', within a cube just big enough to contain the region of
interest. Either choice approximates a uniform mass distribution very
accurately.  Outside the cube we lay down particles on a set of
concentric cubic shells, centered on the cube, which extend outwards
until they fill the entire periodic volume of the parent simulation.
These concentric shells are filled with more massive particles whose
interparticle separation increases approximately linearly with
distance from the high resolution region. Unlike the grid or glass,
this arrangement does not reproduce a perfectly uniform mass
distribution. However, by populating each shell with regularly spaced
particles, we obtain a configuration which is uniform enough for our
purposes.

In the interest of efficiency, we replace those particles in the high-resolution
cube that do not end up in the selected halo with more massive particles made by
combining several high-resolution ones. This procedure, particularly for the
dwarf halo resimulations, significantly reduces the number of particles in the
initial conditions and the run time of the subsequent simulation. Thus, each
halo forms from an ``amoeba-shaped'' region consisting only of the highest
resolution particles in the hierarchy.  We have explicitly checked that the
resampling procedure adds no extra power; in tests, the multi-mass particle
distribution remains very close to uniform over an expansion factor of up to
$\sim 50$.  

Once a multi-mass but uniform mass distribution has been created, the
next step is to to add the appropriate Gaussian density
fluctuations. This is done by assigning a displacement and a peculiar
velocity to each particle using Fourier methods. By using the same
amplitude and phase for every Fourier mode present in the parent
simulation, a perturbed density field essentially identical to that of
the parent simulation can be reproduced.  In the high-resolution cube,
because the particle mass is smaller than in the parent simulation, it
is necessary to add additional short wavelength modes (with amplitudes
fixed by the adopted power spectrum) down to the Nyquist wavelength of
the new particle grid. To ensure that the Fourier transforms needed to
add this extra power are of a manageable size, we make the additional
power periodic on the scale of the central cube rather than on the
scale of the parent simulation. The longest wavelength added is
typically smaller than one tenth of the side length of the original
cube. As described in P03, the individual components of the
displacement field are generated in turn, and the displacements
calculated at the particle positions by trilinear interpolation. To
set up growing modes, we use the Zel'dovich approximation and make the
peculiar velocities proportional to the displacements.

The initial redshift, $z_i$, of each resimulation is chosen so that
density fluctuations in the high-resolution region are in the linear
regime.  P03 find that convergent results are obtained when $z_i$ is
high enough that the (theoretical) rms mass fluctuation on the
smallest resolved mass scale, $\sigma(m_p,z_i)$ does not exceed $\sim
0.3$ (where $m_p$ is the mass of a high-resolution particle). All of
our simulations satisfy this criterion.

\subsection{Halo selection}
\label{ssec:haloselec}

The resimulated halos analyzed in this paper were all identified in
the parent simulations by applying the friends-of-friends (FoF) group
finding algorithm (Davis et al 1985) with a linking length
$l=0.164$. Cluster-sized halos were drawn from a $479\, h^{-1}$ Mpc
simulation volume ($\Lambda$CDM-512 in Table~\ref{tab:numexp}). The
FoF($0.164$) groups were first ordered by mass and then ten
consecutive entries on the list centered around a mass of $10^{15}\,
h^{-1} \, M_{\odot}$ were selected.  Galaxy halos were likewise drawn
from a $35.325\, h^{-1}$ Mpc volume (SGIF-128), with the exception of
three of the halos (G1-G3) which were selected from a $32.5\, h^{-1}$
Mpc volume (ENS01).

Target dwarf halos were also found in the SGIF-128 simulation.
However, because of their extremely low mass (corresponding to 5-6
particles in SGIF-128), it was necessary to create a second ``parent''
simulation for them by resimulating a region of the SGIF-128 volume at
significantly higher resolution.  To this end, a spherical region of
radius $4.4 \, h^{-1}$ Mpc, with mean density close to the universal
average, was selected at random within the $35.325 \, h^{-1}$ Mpc box.
This spherical region was then resimulated with roughly one hundred
times more particles than in SGIF-128. The target dwarf halos were
identified within this spherical volume again from an FoF(0.2) group
list. A total of eighteen halos with 450-550 particles (corresponding
to masses of $9$-$11\times 10^{9}\, h^{-1}\, M_\odot$) were chosen. We
report results on the four halos in this list that have been
resimulated to date. High resolution initial conditions for these
dwarf halos were created in an identical way to the more massive
galaxy and cluster halos.

Numerical parameters were chosen to ensure that all halos, regardless of mass,
were resimulated at comparable mass resolution (typically over $10^6$ particles
within the virial radius at $z=0$, see Table~\ref{tab:halpar}).

\subsection{The Analysis}
\label{ssec:anal}

We focus our analysis on the spherically-averaged mass profile of simulated
halos at $z=0$. This is measured by sorting particles in distance from the
center of each halo and arranging them in bins of equal logarithmic width in
radius. Density profiles, $\rho(r)$, are computed simply by dividing the mass in
each bin by its volume.  The cumulative mass within each bin, $M(r)$, is then
used to compute the circular velocity profile of each halo,
$V_c(r)=\sqrt{GM(r)/r}$, as well as the cumulative density profile, ${\bar
\rho}(r)=3\, M(r)/4\pi r^3$, which we shall use in our analysis.

The center of each halo is determined using an iterative technique in
which the center of mass of particles within a shrinking sphere is
computed recursively until a few thousand particles are left (see P03
for details). In a multi-component system, such as a dark halo with
substructure, this centering procedure isolates the densest region
within the largest subcomponent. In more regular systems, the center
so obtained approximately coincides with the centers defined by the
center of mass weighted by the local density or by the gravitational
potential of each particle.

We note that, unlike in NFW, no attempt has been made to select halos
at a particularly quiet stage in their dynamical evolution; our sample
thus contains halos in equilibrium as well as a few with prominent
substructure as a result of recent accretion events.

\subsection{Parameter selection criteria}
\label{ssec:convcrit}

The analysis presented in P03 demonstrated that the mass profile of a
simulated halo is numerically robust down to a ``convergence radius'',
$r_{\rm conv}$, that depends primarily on the number of particles and
time steps, as well as on the choice of gravitational softening in the
simulation.  Each of these choices imposes a minimum radius for
convergence, although for an ``optimal'' choice of parameters (i.e.,
one that, for given $r_{\rm conv}$, minimizes the number of force
computations and time steps) the most stringent criterion is that
imposed by the number of particles within $r_{200}$.  In this optimal
case, the minimum resolved radius is well approximated by the location
at which the two-body relaxation time, $t_{\rm relax}$, equals the age
of the universe (see Hayashi et al 2003 for further validation of
these criteria, but see Binney 2003 for a different opinion).

To be precise, we shall identify $r_{\rm conv}$ with the radius where $t_{\rm
relax}$ equals the circular orbital timescale at the virial radius, $t_{\rm
circ}(r_{200})=2\pi\, r_{200}/V_{200}$. Thus, $r_{\rm
conv}$ is defined by the following equation,
\begin{equation}
{t_{\rm relax}(r)\over t_{\rm circ}(r_{200})}={N \over 8\, \ln N} {r/V_c \over
r_{200}/V_{200}}=1.
\label{eq:trelax}
\end{equation}
Here $N=N(r)$ is the number of particles enclosed within $r$, and
$V_{200}=V_c(r_{200})$.  With this definition, the convergence radius in our
best-resolved halos, outside which $V_c(r)$ converges to better than $10\%$, is
of order $\sim 0.005\, r_{200}$.

\section{Results}
\label{sec:res}

\subsection{Density Profiles}
\label{ssec:dprof}

\begin{figure*}
\begin{center}
\epsscale{1.5}
\psfig{file=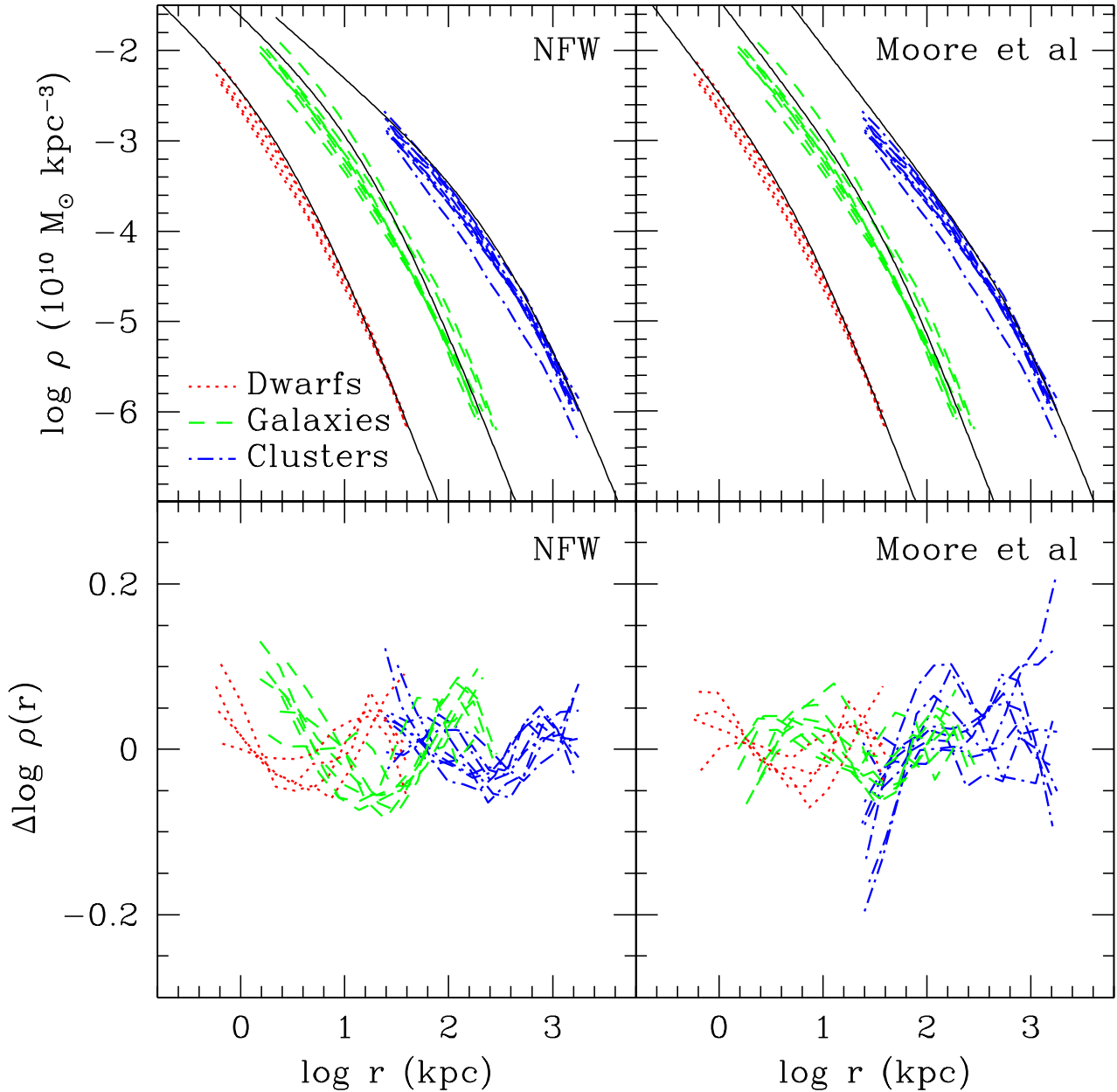,width=180mm}
\end{center}
\caption{
Spherically-averaged density profiles of all our simulated halos. Densities are
computed in radial bins of equal logarithmic width and are shown from the
innermost converged radius ($r_{\rm conv}$) out to about the virial radius of
each halo ($r_{200}$). Our simulations target halos in three distinct mass
groups: ``dwarf'', ``galaxy'', and ``cluster'' halos. These groups span more
than five decades in mass. Thick solid lines in the top panels illustrate the
expected halo profile for each mass range according to the fitting formula
proposed by NFW (top-left) or M99 (top-right). Bottom panels indicate the
deviation from the {\it best} fit achieved for each individual halo (simulation
minus fit) with the NFW profile (eq.~\ref{eq:nfw}) or with its modified form, as
proposed by M99 (eq.~\ref{eq:mooreetal}).
\label{figs:rhoprof}}
\end{figure*}

The top panels of Figure~\ref{figs:rhoprof} show the density profiles,
$\rho(r)$, of the nineteen simulated halos in our sample. In physical units, the
profiles split naturally into three groups: from left to right, ``dwarf''
(dotted), ``galaxy'' (dashed), and ``cluster'' (dot-dashed) halos,
respectively. Each profile is shown from the virial radius, $r_{200}$, down to
the innermost converged radius, $r_{\rm conv}$; a convention that we shall adopt
in all figures throughout this paper.

The thick solid lines in the top-left panel show the NFW profiles
(eq.~\ref{eq:nfw}) expected for halos in each group, with parameters chosen
according to the prescription of Eke, Navarro \& Steinmetz (2001). Note that
these NFW curves are {\it not} best fits to any of the simulations, but that
they still capture reasonably well the shape and normalization of the density
profiles of the simulated halos.

The top right panel of Figure~\ref{figs:rhoprof} is similar to the top left one,
but the comparison is made here with the modified form of the NFW profile
proposed by M99 (eq.~\ref{eq:mooreetal}). There is no published prescription
specifying how to compute the numerical parameters of this formula for halos of
given mass, so the three profiles shown in this panel are just ``eyeball'' fits
to one halo in each group. Like the NFW profile, the M99 formula also appears to
describe reasonably well the gently-curving density profiles of $\Lambda$CDM
halos.

Figure~\ref{figs:rhoprof} thus confirms a number of important trends that were
already evident in prior simulation work.

\begin{itemize}

\item 
$\Lambda$CDM halo density profiles deviate significantly from simple power laws,
and steepen systematically from the center outwards; they are shallower than
isothermal near the center and steeper than isothermal near the virial radius.

\item
There is no indication of a well defined central ``core'' of constant density;
the dark matter density keeps increasing all the way in, down to the innermost
resolved radius.

\item
Simple formulae such as the NFW profile (eq.~\ref{eq:nfw}) or the M99 formula
(eq.~\ref{eq:mooreetal}) appear to describe the mass profile of all halos
reasonably well, irrespective of mass, signaling a ``universal'' profile
shape. Properly scaled, a dwarf galaxy halo is almost indistinguishable from a
galaxy cluster halo.

\end{itemize}

We elaborate further on each of these conclusions in what follows.

\subsubsection{NFW vs M99 fits}
\label{sssec:nfwvsm98}

Are the density profiles of $\Lambda$CDM halos described better by the
NFW formula (eq.~\ref{eq:nfw}) or by the modification proposed by M99
(eq.~\ref{eq:mooreetal})? The answer may be seen in the bottom panels
of Figure~\ref{figs:rhoprof}. These panels show the deviations
(simulation minus fit) from the {\it best} fits to the density
profiles of each halo using the NFW profile or the M99 profile. These
fits are obtained by straightforward $\chi^2$ minimization, assigning
equal weight to each radial bin. This is done because the statistical
(Poisson) uncertainty in the determination of the density within each
bin is negligible (each bin contains from several thousand to several
hundred thousand particles) so the remaining uncertainties are likely
to be dominated by systematics, such as the presence of substructure,
varying asphericity, as well as numerical error, whose radial
dependence is difficult to assess quantitatively (see P03).

As shown in the bottom panels of Figure~\ref{figs:rhoprof}, there is significant
variation in the shape of the density profile from one halo to another. Some
systems are fit better by eq.~\ref{eq:nfw} than by eq.~\ref{eq:mooreetal}, and
the reverse is true in other cases.  Over the radial range resolved by the
simulations, $\rho(r)$ deviates from the best fits by less than $\sim 50\%$. NFW
fits tend to {\it underestimate} the density in the inner regions of most halos;
by up to $35\%$ at the innermost resolved point. M99 fits, on the other hand,
seem to do better for low mass halos, but tend to {\it overestimate} the density
in the inner regions of cluster halos by up to $60\%$. We have explicitly
checked that these conclusions are robust to reasonable variation in the binning
used to construct the density profiles, as well as in the adopted minimization
procedure.

This level of accuracy may suffice for a number of observational
applications, with the proviso that comparisons are restricted to
radii where numerical simulations are reliable; i.e., $r_{\rm conv} <
r <r_{200}$. Deviations from the best fits increase systematically
towards the center, so it is likely that extrapolations of either
fitting formula to radii much smaller than $r_{\rm conv}$ will incur
substantial error. We discuss below (\S~\ref{ssec:newprof}) possible
modifications to the fitting formulae that may minimize the error
introduced by these extrapolations.

\subsection{Circular Velocity Profiles}
\label{ssec:vcprof}

Many observations, such as disk galaxy rotation curves or strong gravitational
lensing, are better probes of the {\it cumulative} mass distribution than
of the differential density profile shown in Figure~\ref{figs:rhoprof}. Since
cumulative profiles are subject to different uncertainties than differential
ones, it is important to verify that our conclusions regarding the suitability
of the NFW or M99 fitting formulae are also applicable to the cumulative mass
distribution of $\Lambda$CDM halos.

\begin{figure*}
\begin{center}
\epsscale{1.5}
\psfig{file=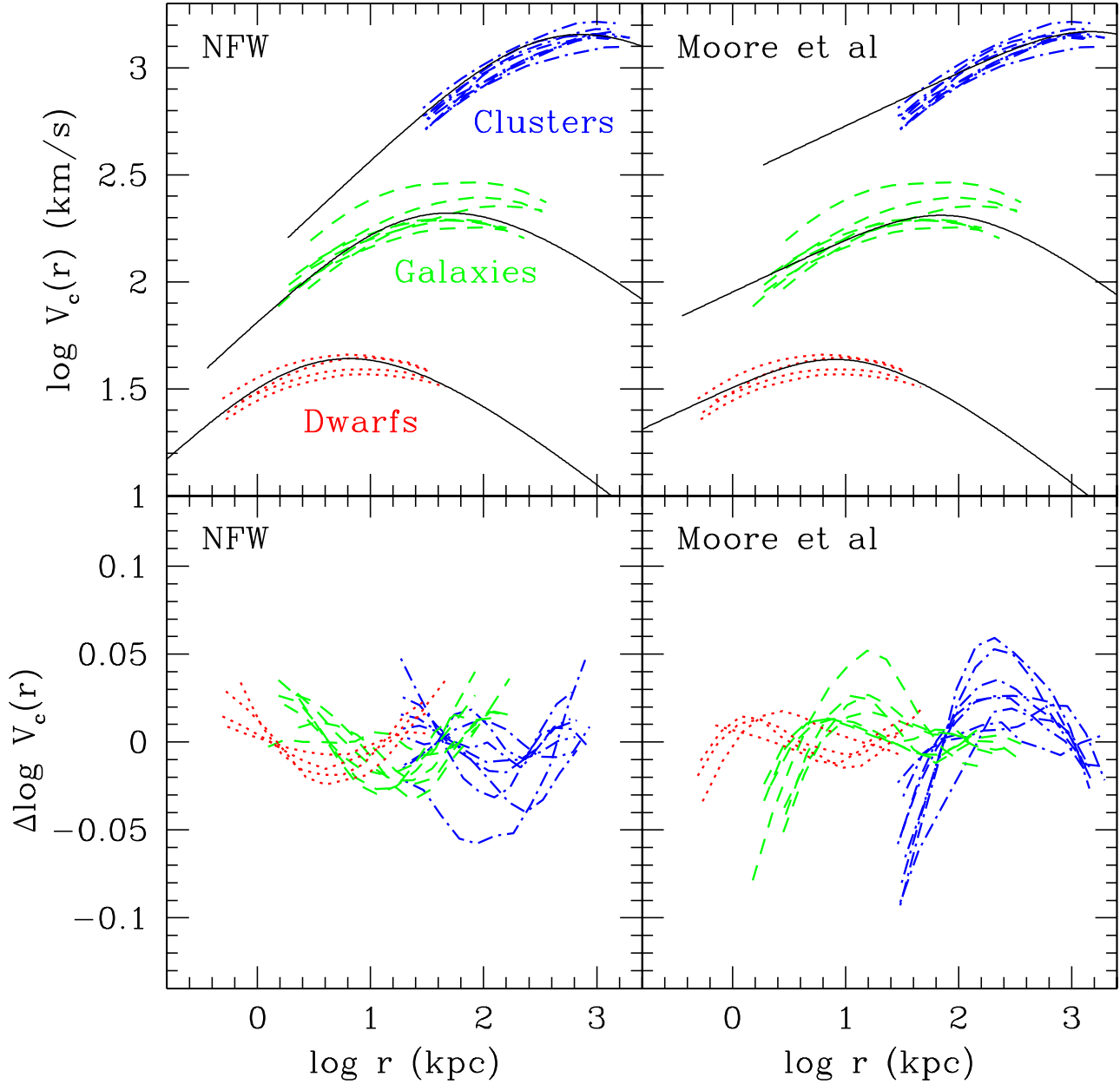,width=180mm}
\end{center}
\caption{
Spherically-averaged circular velocity ($V_c(r)=\sqrt{GM(r)/r}$) profiles of all
our simulated halos. As in Figure~\ref{figs:rhoprof}, circular velocities are
computed in radial bins of equal logarithmic width and are shown from the
innermost converged radius ($r_{\rm conv}$) out to about the virial radius
($r_{200}$) of each halo. Our simulations target halos in three distinct mass
groups: ``dwarf'', ``galaxy'', and ``cluster'' halos, spanning more than a
factor of $\sim 50$ in velocity.  Thick solid lines in the top panels illustrate
the expected profile for each mass range according to the fitting formula
proposed by NFW (top-left) or M99 (top-right). Bottom panels indicate the
deviation from the {\it best} fit achieved for each individual halo (simulation
minus fit) with the NFW profile (eq.~\ref{eq:nfw}) or with its modified form, as
proposed by M99 (eq.~\ref{eq:mooreetal}).
\label{figs:vcprof}}
\end{figure*}

The radial dependence of the spherically-averaged circular velocity profile of
all halos in our series is shown in Figure~\ref{figs:vcprof}. As in
Figure~\ref{figs:rhoprof}, the thick solid curves in the top left (right) panel
are meant to illustrate a typical NFW (M99) profile corresponding to dwarf,
galaxy, and cluster halos, respectively.  The bottom left and right panels show
deviations from the {\it best} fit to each halo using the NFW or M99 profile,
respectively.  Both profiles reproduce the cumulative mass profile of the
simulated halos reasonably well. The largest deviations seen are for the M99
fits, but they do not exceed $25\%$ over the radial range resolved in the
simulations. NFW fits fare better, with deviations that do not exceed $10\%$.

As with the density profiles, the deviations between simulation and fits,
although small, increase toward the center, suggesting that caution should be
exercised when extrapolating these fitting formulae beyond the spatial region
where they have been validated. This is important because observational data,
such as disk galaxy rotation curves, often extend to regions inside the minimum
convergence radius in these simulations.

\subsection{Radial dependence of logarithmic slopes}
\label{ssec:logslope}

\begin{figure*}
\begin{center}
\epsscale{1.25}
\psfig{file=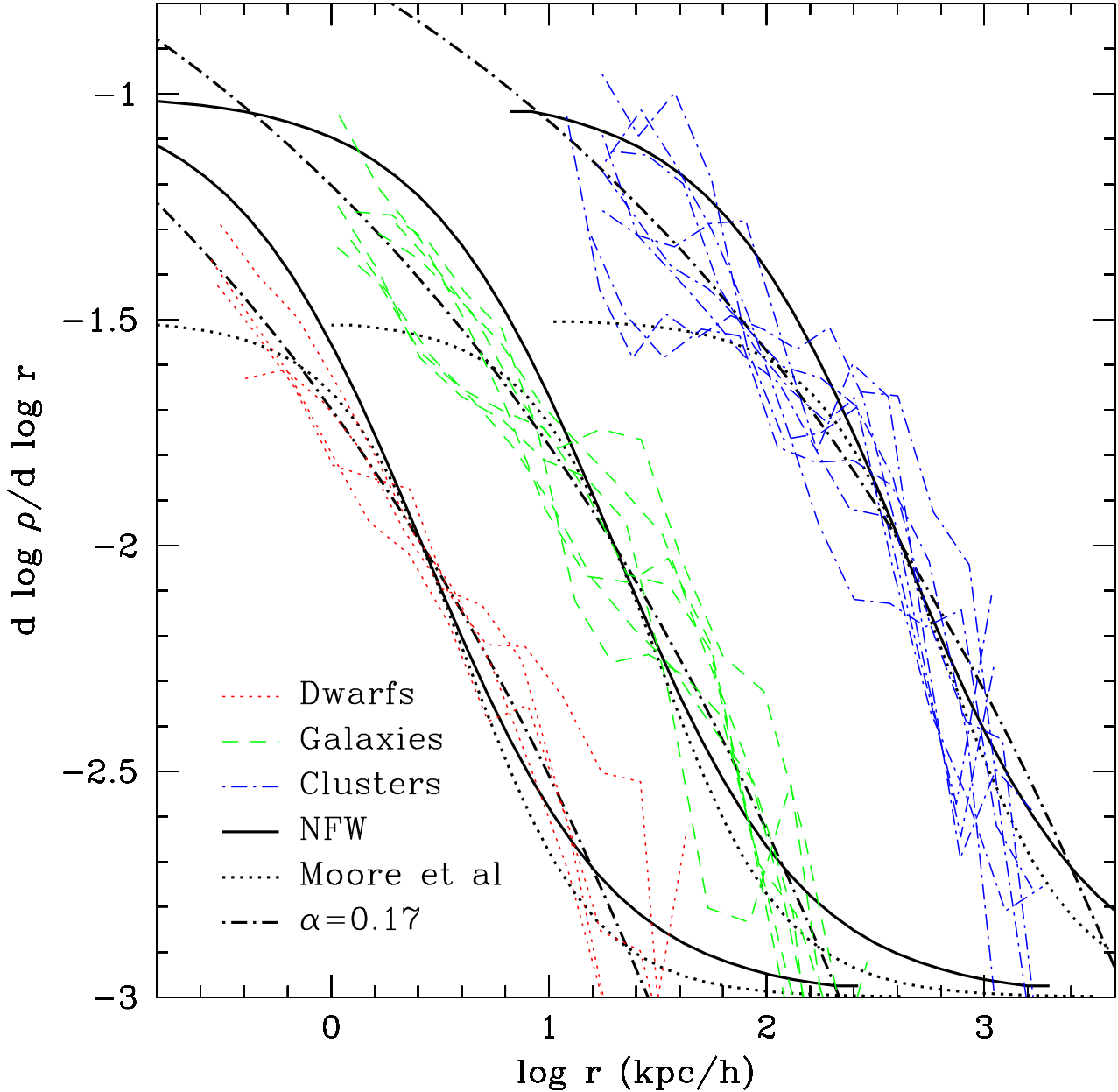, width=180mm}
\end{center}
\caption{Logarithmic slope of the density profile of all halos in our sample,
plotted versus radius. Thick solid and dotted curves illustrate the radial
dependence of the slope expected from the NFW profile (eq.~\ref{eq:nfw}) and the
modification proposed by M99 (eq.~\ref{eq:mooreetal}), respectively. Note that
although both fitting formulae have well-defined asymptotic inner slopes ($-1$
and $-1.5$, respectively) there is no sign of convergence to a well-defined
value of the central slope in the simulated halos. At the innermost converged
radius, the simulated halo profiles are shallower than $-1.5$, in disagreement
with the Moore et al profile. Also, inside the radius at which the slope equals
$-2$, $r_{-2}$, the profiles appear to get shallower more gradually than in the
NFW formula. A power-law radial dependence of the slope seems to fit the results
of our simulations better; the dot-dashed lines indicate the predictions of the
$\rho_{\alpha}$ profile introduced in eqs.~\ref{eq:newbeta} and
~\ref{eq:newprof} for $\alpha=0.17$. Best fits to individual halos yield
$\alpha$ in the range $0.1$-$0.2$ (see Table~\ref{tab:fitpar}).
\label{figs:slopes}}
\end{figure*}

We have noted in the previous subsections that systematic deviations are
noticeable in both NFW and M99 fits to the mass profiles of simulated
$\Lambda$CDM halos. NFW fits tend to underestimate the dark matter density near
the center, whilst M99 fits tend to overestimate the circular velocity in the
inner regions. The reason for this is that {\it neither} fitting formula fully
captures the radial dependence of the density profile.  We explore this in
Figure~\ref{figs:slopes}, which shows the logarithmic slope, $d\ln\rho/d\ln r
\equiv -\beta(r)$, of all simulated halos, as a function of radius. Although
there is substantial scatter from halo to halo, a number of trends are robustly
defined.

The first trend to note is that halo density profiles become shallower inward
down to the innermost resolved radius, $r_{\rm conv}$ (the smallest radius
plotted in Figure~\ref{figs:slopes}). {\it We see no indication for convergence
to a well defined asymptotic value of the inner slope in our simulated halos},
neither to the $\beta_0=\beta(r=0)=1$ expected for the NFW profile (solid curves
in Figure~\ref{figs:slopes}) nor to the $\beta_0=1.5$ expected in the case of
M99 (dotted curves in same figure).

The second trend is that the radial dependence of the logarithmic slope deviates
from what is expected from either the NFW or the M99 fitting formulae. Near
$r_{\rm conv}$, the slopes are significantly shallower than $\beta_0=1.5$ (and
thus in disagreement with the M99 formula) but they are also significantly
steeper than expected from NFW fits. In quantitative terms, let us consider the
slope well inside the characteristic radius, $r_{-2}$ (where the slope takes the
``isothermal'' value
\footnote{The characteristic radius, $r_{-2}$, as well as the density at that
radius, $\rho_{-2}\equiv \rho(r_{-2}$, can be measured directly from the
simulations, without reference to or need for any particular fitting
formula. For the NFW profile, $r_{-2}$ is equivalent to the scale radius $r_s$
(see eq.~\ref{eq:nfw}). The density at $r_{-2}$ is related to the NFW
characteristic density, $\rho_s$, by $\rho_{-2}\equiv \rho(r_{-2})=\rho_s/4$.}
of $\beta(r_{-2})=2$).  For cluster halos, for example, at $r=0.1\, r_{-2}$
($\sim 50 \, h^{-1}$ kpc) the average slope is approximately $-1.3$, whereas the
NFW formula predicts $\sim -1.18$ and M99 predicts $\sim -1.5$. This is in
agreement with the latest results of Fukushige, Kawai \& Makino (2003), who also
report profiles shallower than $r^{-1.5}$ at the innermost converged radius of
their simulations. A best-fit slope of $r^{-1.3}$ was also reported by Moore et
al (2001) for a dwarf galaxy halo (of mass similar to the Draco dwarf
spheroidal), although that simulation was stopped at $z=4$, and might therefore
not be directly comparable to the results we present here.

This discrepancy in the radial dependence of the logarithmic slope
between simulations and fitting formulae is at the root of the
different interpretations of the structure of the central density cusp
proposed in the literature. For example, because profiles become
shallower inward more gradually than in the NFW formula, modifications
with more steeply divergent cusps (such as eq.~\ref{eq:mooreetal})
tend to fit density profiles (but not circular velocity profiles)
better in the region interior to $r_{-2}$. This is {\it not}, however,
a sure indication of a steeper cusp. Indeed, {\it any} modification to
the NFW profile that results in a more gradual change in the slope
inside $r_{-2}$ will lead to improved fits, {\it regardless} of the
value of the asymptotic central slope. We show this explicitly below
in \S~\ref{ssec:newprof}.

\subsection{Maximum asymptotic slope}
\label{ssec:asyslope}

Conclusive proof that the central density cannot diverge as steeply as
$\beta_0=1.5$ is provided by the total mass inside the innermost resolved
radius, $r_{\rm conv}$. This is because, at any radius $r$, the mean density,
$\bar \rho(r)$, together with the local density, $\rho(r)$, provide a robust
upper limit to the asymptotic inner slope. This is given by $\beta_{\rm
max}(r)=3(1-\rho(r)/{\bar \rho(r)}) > \beta_0$, under the plausible assumption
that $\beta$ is monotonic with radius.

\begin{figure*}
\begin{center}
\epsscale{1.25}
\psfig{file=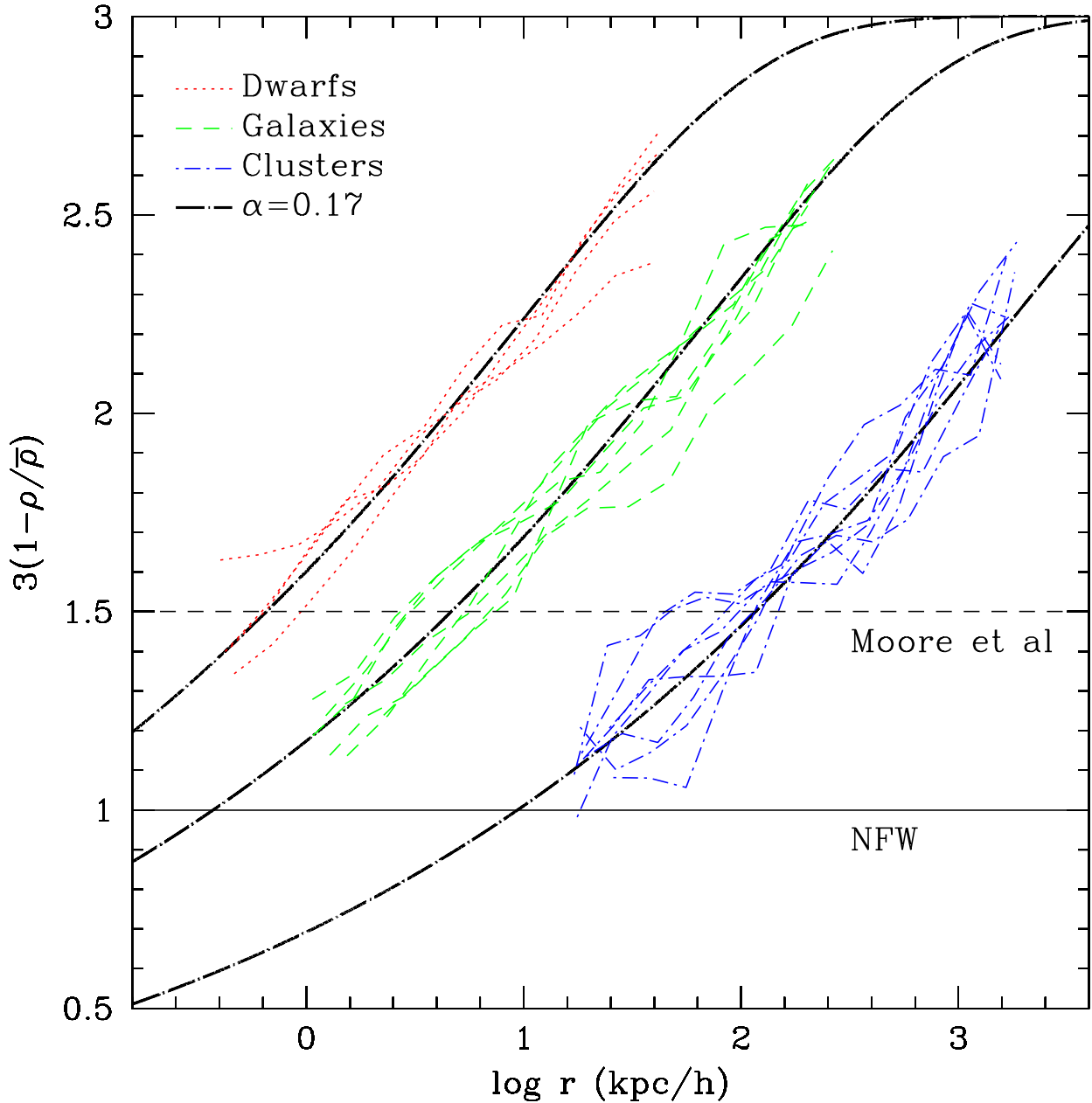, width=180mm}
\end{center}
\caption{Maximum asymptotic inner slope compatible with the mean density
interior to radius $r$, $\bar \rho(r)$, and with the local density at that
radius, $\rho(r)$. This provides a robust limit to the central slope,
$\beta_0<\beta_{\rm max}(r)=-3(1-\rho(r)/{\bar \rho(r)})$, under the plausible
assumption that $\beta$ is monotonic with radius. Note that there is not enough
mass within the innermost converged radius in our simulations to support density
cusps as steep as $r^{-1.5}$. The asymptotic slope of the NFW profile,
$\beta_0=1$, is still compatible with the simulated halos, although there is no
convincing evidence for convergence to a well defined power-law behavior in any
of our simulated halos. The thick dot-dashed curves illustrate the expected
radial dependence of $\beta_{\rm max}$ for the $\rho_{\alpha}$ profile
introduced in \S~\ref{ssec:newprof}, for $\alpha=0.17$.
\label{figs:betaprof}}
\end{figure*}

Figure~\ref{figs:betaprof} shows $\beta_{\rm max}$ as a function of radius;
clearly, except for possibly one dwarf system, no simulated halo has enough dark
mass within $r_{\rm conv}$ to support cusps as steep as $r^{-1.5}$. The NFW
asymptotic slope, corresponding to $\beta_0=1$, is still consistent with the
simulation data, but the actual central value of the slope may very well be
shallower.  We emphasize again that there is no indication for convergence to a
well defined value of $\beta_0$: density profiles become shallower inward down
to the smallest resolved radius in the simulations.

\subsection{A ``universal'' density profile}

Figure~\ref{figs:slopes} shows also that there is a well-defined trend with
mass in the slope of the density profile measured at $r_{\rm conv}\sim 0.005$
to $0.01r_{200}$ (the innermost point plotted for each profile): $\beta(r_{\rm conv})
\sim 1.1$ for clusters, $\sim 1.2$ for galaxies, and $\sim 1.35$ for dwarfs. A
similar trend was noted by Jing \& Suto (2000), who used it to argue against a
``universal'' density profile shape. However, as discussed by Klypin et al
(2001), this is just a reflection of the trend between the concentration of a
halo and its mass. It does {\it not} indicate any departure from similarity in
the profile shape. Indeed, one does {\it not} expect the profiles of halos of
widely different mass, such as those in our series, to have similar slopes at a
constant fraction of the virial radius. Rather, if the density profiles are
truly self-similar, slopes ought to coincide at fixed fractions of a {\it
mass-independent} radial scale, such as $r_{-2}$.

\begin{figure*}
\begin{center}
\epsscale{2.5}
\centerline{\epsfig{figure=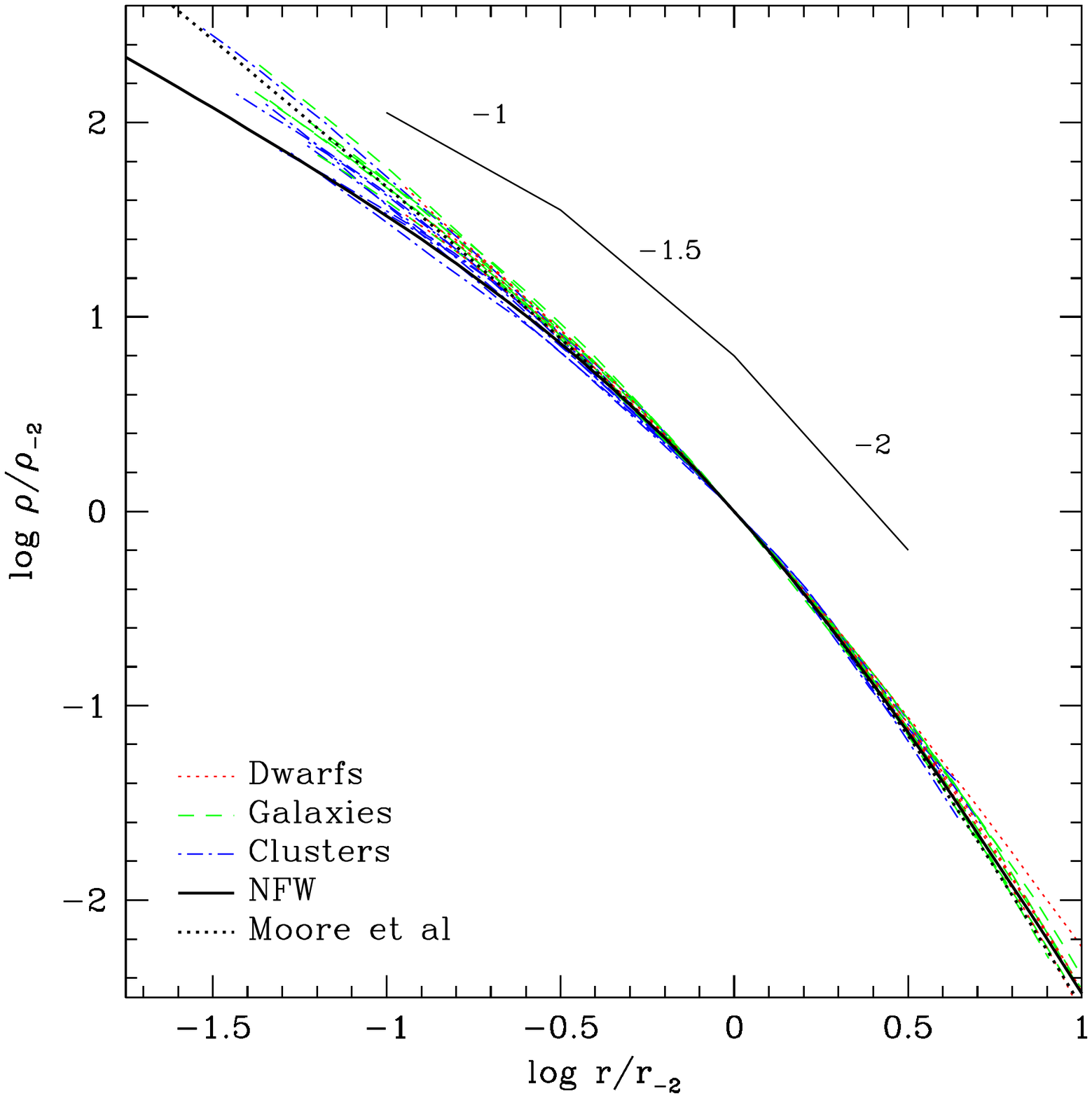,width=0.575\linewidth}}
\centerline{\epsfig{figure=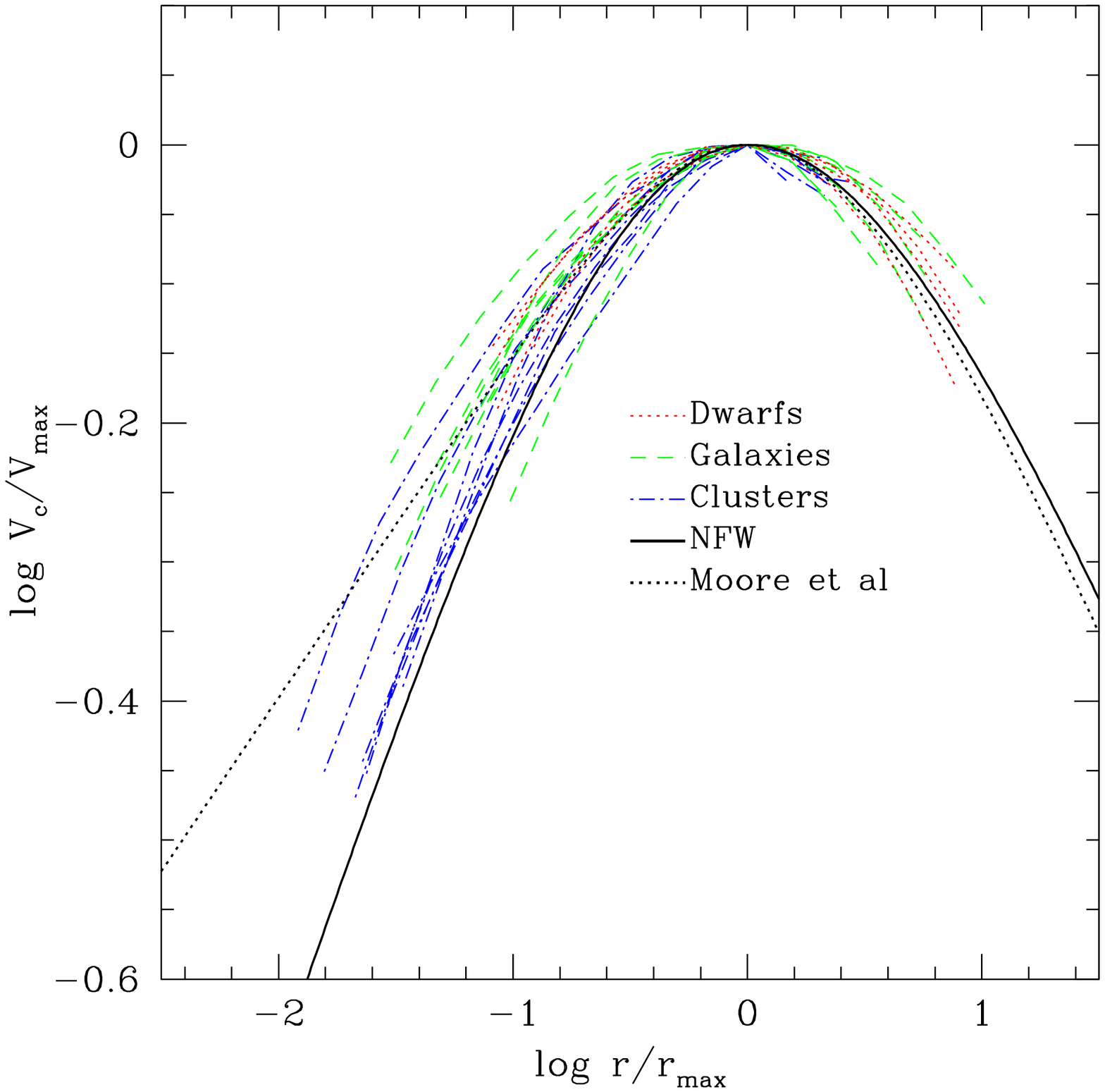,width=0.575\linewidth}}
\end{center}
\caption{(a-left) Density profiles of all halos in our series, scaled to the
radius, $r_{-2}$, where the local logarithmic slope of the density profile takes
the isothermal value of $\beta=-d\log\rho/d\log r=2$. Densities are scaled to
$\rho_{-2}=\rho(r_{-2})$. This figure shows that, with proper scaling, there is
little difference in the shape of the density profile of halos of different
mass, confirming the ``universal'' nature of the mass profile of $\Lambda$CDM
halos. The NFW profile (eq.~\ref{eq:nfw}) is a fixed curve in these scaled
units, and is shown with a thick solid line. The M99 formula
(eq.~\ref{eq:mooreetal}) is shown with a dashed line. (b-right) Circular
velocity profiles all halos in our series, scaled to the maximum velocity,
$V_{\rm max}$, and to the radius at which it is reached, $r_{\rm max}$. Note the
significant scatter from halo to halo, and also that the NFW and M99 profiles
appear to bracket the extremes of the mass profile shapes of halos in our
simulation series.
\label{figs:rhoprofsc}}
\end{figure*}

Figure~\ref{figs:rhoprofsc}a shows the striking similarity between the structure
of halos of different mass when all density profiles are scaled to $r_{-2}$ and
$\rho_{-2}\equiv \rho(r_{-2})$.  The density profile of a dwarf galaxy halo then
differs very little from that of a galaxy cluster $10^5$ times more
massive. This demonstrates that spherically-averaged density profiles are
approximately ``universal'' in shape; rarely do individual density profiles
deviate from the scaled average by more than $\sim 50\%$.

In the scaled units of Figure~\ref{figs:rhoprofsc}, the NFW and M99 profiles are
fixed, and are shown as solid and dotted curves, respectively. With this
scaling, differences between density profiles are more evident than when best
fits are compared, since the latter --- by definition --- minimize the
deviations.  In Figure~\ref{figs:rhoprofsc}a, for example, it is easier to
recognize the ``excess'' of dark mass inside $r_{-2}$ relative to the NFW
profile that authors such as M99 and Fukushige \& Makino (1997, 2001, 2003) have
(erroneously) interpreted as implying a steeply divergent density cusp.

The similarity in mass profile shapes is also clear in
Figure~\ref{figs:rhoprofsc}b, which shows the circular velocity curves
of all halos in our series, scaled to the maximum, $V_{\rm max}$ and
to the radius where it is reached, $r_{\rm max}$. NFW and M99 are
again fixed curves in these scaled units. This comparison is more
relevant to observational interpretation, since rotation curve,
stellar dynamical, and lensing tracers are all more directly related
to $V_c(r)$ than to $\rho(r)$. Because of the reduced dynamic range of
the y-axis, the scatter in mass profiles from halo to halo is more
clearly apparent in the $V_c$ profiles; the NFW and M99 profiles
appear to approximately bracket the extremes in the mass profile
shapes of simulated halos. We discuss below a simple fitting formula
that, with the aid of an extra parameter, is able to account for the
variety of mass profile shapes better that either the NFW or M99
formula.

\subsection{An improved fitting formula}
\label{ssec:newprof}

Although the discussion in the previous subsections has concentrated on global
deviations from simple fitting formulae such as NFW or M99, it is important to
emphasize again that such deviations, although significant, are actually rather
small. As shown in Figure~\ref{figs:vcprof}, best NFW fits reproduce the
circular velocity profiles to an accuracy of better than $\sim 10\%$ down to
roughly $0.5\%$ of $r_{200}$. Although this level of accuracy may suffice for
some observational applications, the fact that deviations increase inward and
are maximal at the innermost converged point suggests the desirability of a new
fitting formula better suited for extrapolation to regions beyond those probed
reliably by simulations.

An improved fitting formula ought to reproduce: (i) the more gradual shallowing
of the density profile towards the center; (ii) the apparent lack of evidence
for convergence to a well-defined central power-law; and (iii) the significant
scatter in profile shape from halo to halo. After some experimentation, we have
found that a density profile where $\beta(r)$ is a power-law of radius is a
reasonable compromise that satisfies these constraints whilst retaining
simplicity;
\smallskip
\begin{equation}
\beta_{\alpha}(r)=-{d\ln \rho/d\ln r}=2\left({r/r_{-2}}\right)^{\alpha},
\label{eq:newbeta}
\end{equation}
\smallskip
which corresponds to a density profile of the form,
\smallskip
\begin{equation}
\ln (\rho_{\alpha}/\rho_{-2})={(-2/\alpha)} [\left({r/r_{-2}}\right)^{\alpha} -1].
\label{eq:newprof}
\end{equation}
\smallskip
This profile has finite total mass (the density cuts off exponentially at large
radius) and has a logarithmic slope that decreases inward more gradually than
the NFW or M99 profile.  The thick dot-dashed curves in
Figures~\ref{figs:slopes} and ~\ref{figs:betaprof} show that
eq.~\ref{eq:newprof} (with $\alpha\sim 0.17$) does indeed reproduce fairly well
the radial dependence of $\beta(r)$ and $\beta_{\rm max}(r)$ in simulated halos.

\begin{figure*}
\begin{center}
\epsscale{2.0}
\psfig{file=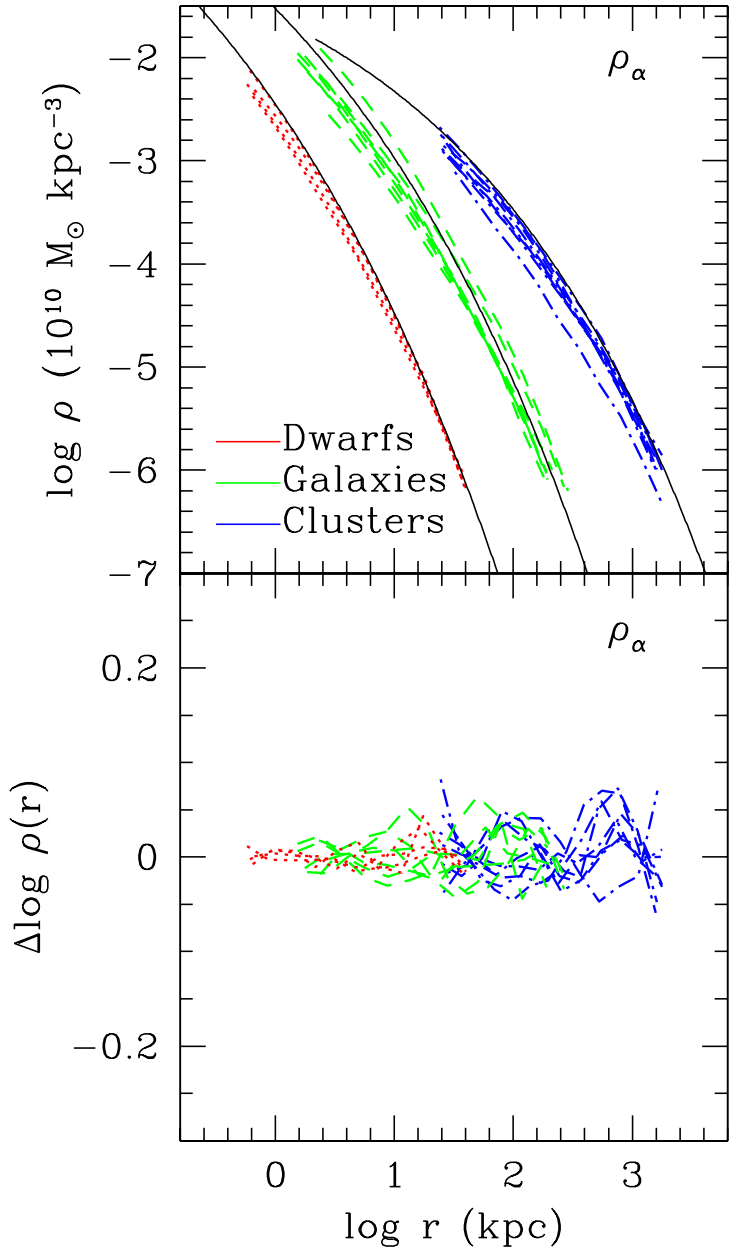,width=180mm}
\end{center}
\caption{
As Figure~\ref{figs:rhoprof}, but for the $\rho_{\alpha}$ fitting formula
presented in eq.~\ref{eq:newprof}.  Thick solid lines in the top panels
illustrate the expected halo profile for each mass range according to the
prescription proposed by NFW. Bottom panels indicate the deviation from the {\it
best} $\rho_{\alpha}$ fit achieved for each individual halo, taking $\alpha$ as
a free parameter. Note the improvement in the fits compared with those achieved
with the NFW or M99 profile and shown in Figure~\ref{figs:rhoprof}.
\label{figs:rhoprof_alpha}}
\end{figure*}

Furthermore, adjusting the parameter $\alpha$ allows the profile to be tailored
to each individual halo, resulting in improved fits. Indeed, as shown in
Figure~\ref{figs:rhoprof_alpha}, eq.~\ref{eq:newprof} reproduces the density
profile of individual halos to better than $\sim 10\%$ over the reliably
resolved radial range, and that there is no discernible radial trend in the
residuals. This is a significant improvement over NFW or M99 fits, where the
maximum deviations were found at the innermost resolved radius. The best-fit
values of $\alpha$ (in the range $0.1$ - $0.2$) show no obvious dependence on
halo mass, and are listed in Table~\ref{tab:fitpar}. The average $\alpha$ is
$0.172$ and the dispersion about the mean is $0.032$.

We note that the $\rho_{\alpha}$ profile is not formally divergent, and
converges to a finite density at the center, $\rho_0= e^{2/\alpha}\, \rho_{-2}
\sim 6 \times 10^5 \rho_{-2}$ (for $\alpha=0.15$). It is unclear at this point
whether such asymptotic behavior is a true property of $\Lambda$CDM halos or
simply an artifact of the fitting formula that results from choosing $\beta_0=0$
in eq.~\ref{eq:newbeta}. The simulations show no evidence for convergence to a
well-defined central value for the density, but even in the best-resolved cases
they only probe regions where densities do not exceed $\sim 10^2 \,
\rho_{-2}$. This is, for $\alpha$ in the range $0.1$ - $0.2$, several orders of
magnitude below the maximum theoretical limit in eq.~\ref{eq:newprof}. 

We note as well that the convergence to $\beta_0=0$ is quite slow for the values
of $\alpha$ favored by our fits. Indeed, for $\alpha=0.1$, the logarithmic slope
only reaches a value significantly shallower than the NFW asymptotic slope at
radii that are well inside the convergence radius of our simulations; for
example, $\beta_{\alpha}(r)$ only reaches $0.5$ at $r=9.5 \times 10^{-7} \,
r_{-2}$, corresponding to $r \sim 0.01$ pc for galaxy-sized halos. This implies
that the $\rho_{\alpha}$ profile is in practice ``cuspy'' for most astrophysical
applications. Establishing conclusively whether $\Lambda$CDM halos actually have
divergent inner density cusps is a task that awaits simulations with much
improved resolution than those presented here.

\subsection{Comparison between fitting formulae}
\label{ssec:compfit}

\begin{figure*}[t]
\begin{center}
\epsscale{2.0}
\psfig{file=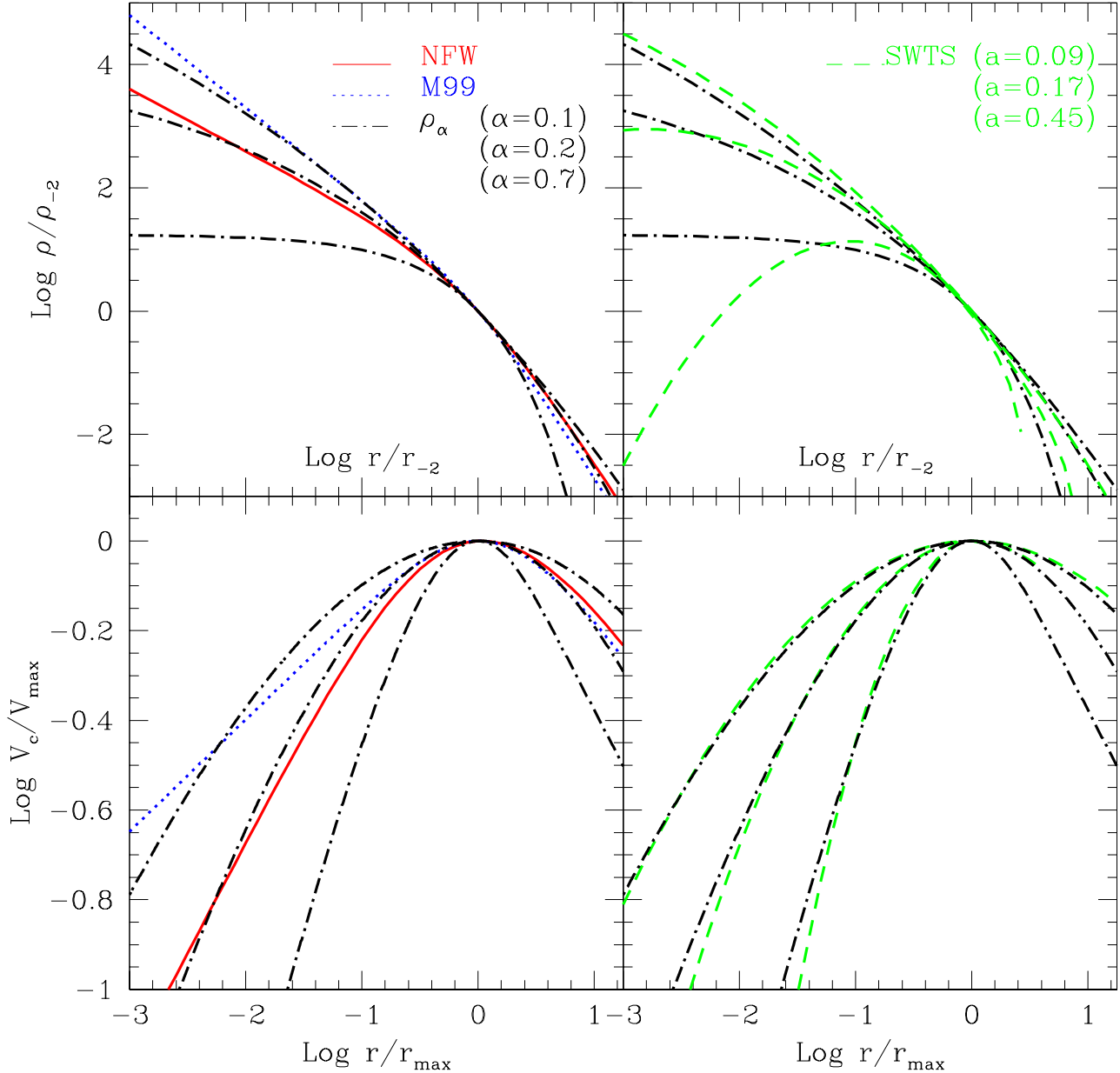, width=180mm}
\end{center}
\caption{Comparison between the density (top) and circular velocity (bottom) profiles
corresponding to four different fitting formulae: NFW (solid curves,
eq.~\ref{eq:nfw}), M99 (dotted curves, eq.~\ref{eq:mooreetal}), SWTS (dashed
curves, eq.~\ref{eq:swts}), and $\rho_{\alpha}$ (dot-dashed curves,
eq.~\ref{eq:newprof}). Circular velocity profiles are scaled to the maximum,
$V_{\rm max}$, and to the radius where that is reached, $r_{\rm max}$. Density
profiles are scaled as in Figure~\ref{figs:rhoprofsc}. Note that, despite having
a finite central density, the $\rho_{\alpha}$ formula matches, for about 3
decades in radius, the NFW profile (for $\alpha=0.2$) or the M99 profile (for
$\alpha=0.1$, see top left panel). It also matches closely the SWTS
``parabolic'' circular velocity profiles intended to reproduce {\it
substructure} halos (see bottom right panel); the $V_c$ profile with
$\alpha=0.7$ is very similar to the SWTS profile with $a=0.45$, the median value
of the fits to substructure halos reported by SWTS. See text for further
discussion.
\label{figs:profcomp}}
\end{figure*}

Figure~\ref{figs:profcomp} compares the density and circular velocity profiles
implied by the $\rho_{\alpha}$ formula (eq.~\ref{eq:newprof}) with the NFW and
M99 profiles (left panels), as well as with the fitting formula proposed by
Stoehr et al (2002, hereafter SWTS) to describe the structure of {\it
substructure} halos (right panels).

The top left panel of Figure~\ref{figs:profcomp} shows that, despite its finite
central density, the $\rho_{\alpha}$ profile can approximate fairly well both an
NFW profile (for $\alpha \sim 0.2$) and an M99 profile (for $\alpha\sim 0.1$)
for over three decades in radius. The circular velocity profile for $\alpha=0.2$
is likewise quite similar to NFW's (bottom left panel of
Figure~\ref{figs:profcomp}), but the similarity to the shape of the M99 $V_c$
profile is less for all values of $\alpha$.

Interestingly, the $V_c$ profiles corresponding to $\rho_{\alpha}$ resemble
parabolae in a $\log$-$\log$ plot, and thus may be used to approximate as well
the mass profiles of substructure halos, as discussed by SWTS. This is
demonstrated in the bottom right panel of Figure~\ref{figs:profcomp}, where we
show that the $V_c$ profiles corresponding to $\alpha=0.1$, $0.2$, and $0.7$,
are very well approximated by the SWTS formula, 
\smallskip
\begin{equation}
\log(V_c/V_{\rm max})=-a[\log(r/r_{\rm max})]^2,
\label{eq:swts}
\end{equation}
\smallskip
for $a=0.09$, $0.17$, and $0.45$, respectively.  The latter value ($a=0.45$, or
$\alpha=0.7$) corresponds to the median of the best SWTS fits to the mass
profile of substructure halos. Note that this is quite different from the
$\alpha\sim 0.1$ - $0.2$ required to fit isolated $\Lambda$CDM halos (see
Table~\ref{tab:fitpar}).

It might actually be preferable to adopt the $\rho_{\alpha}$ profile rather than
the SWTS formula for describing substructure halos, since $\rho_{\alpha}(r)$ is
monotonic with radius and extends over all space. This is not the case for SWTS,
as shown in the top-right panel of Figure~\ref{figs:profcomp}. The SWTS density
profiles are ``hollow'' (i.e., the density has a minimum at the center), and
extend out to a maximum radius, given by $e^{1/4a}\, r_{\rm max}$.  This is
because the circular velocity in the outer regions of the SWTS formula fall off
faster than Keplerian, and therefore the corresponding density becomes
formally negative at a finite radius.

The $\rho_{\alpha}$ profile thus appears versatile enough to reproduce, with a
single fitting parameter, the structure of $\Lambda$CDM halos and of their
substructure. Since $\rho_{\alpha}$ captures the inner slopes better than either
the NFW or M99 profile, it is also likely to be a safer choice should
extrapolation of the mass profile beyond the converged radius prove
necessary. We end by emphasizing, however, that all simple fitting formulae have
shortcomings, and that {\it direct comparison with simulations rather than with
fitting formulae should be attempted whenever possible}.

\subsection{Scaling parameters}
\label{ssec:scpar}

The application of fitting formulae such as the one described above requires a
procedure for calculating the characteristic scaling parameters for a given halo
mass, once the power spectrum and cosmological parameters are specified. NFW
developed a simple procedure for calculating the parameters corresponding to
halos of a given mass. Because of the close relationship between the scale
radius, $r_s$, and characteristic density, $\rho_s$, of the NFW profile and the
$r_{-2}$ and $\rho_{-2}$ parameters of eq.~\ref{eq:newprof}, we can use the
formalism developed by NFW to compute the expected values of these parameters in
a given cosmological model.

NFW interpreted the characteristic density of a halo as reflecting the density
of the universe at a suitably defined time of collapse.  Their formalism assigns
to each halo of mass $M$ (identified at $z=0$) a collapse redshift, $z_{\rm
coll}(M,f)$ defined as the epoch when half the mass of the halo was first
contained in progenitors more massive than a certain fraction $f$ of the final
mass. With this definition, and once $f$ has been chosen, $z_{\rm coll}$ can be
computed using the Press-Schechter theory (e.g., Lacey \& Cole 1993). The NFW
model then assumes that the characteristic density of a halo (i.e., $\rho_s$ in
eq.~\ref{eq:nfw}) is proportional to the mean density of the universe at $z_{\rm
coll}$.

The redshift dependence of the characteristic density was first probed in detail
by Bullock et al (2001, hereafter B01), who proposed a modification to NFW's
model in which, for a given halo mass, the scale radius, $r_s$, remains approximately
constant with redshift.  Eke, Navarro \& Steinmetz (2001, hereafter ENS), on the
other hand, argued that the characteristic density of a halo is determined by the
{\it amplitude and shape} of the power spectrum, as well as by the universal
expansion history. Their formalism reproduces nicely the original results of NFW
as well as the redshift dependence pointed out by B01, and is applicable to more
general forms of the power spectrum, including the ``truncated'' power spectra
expected in scenarios such as warm dark matter (see ENS for more details).

\begin{figure*}[t]
\begin{center}
\epsscale{1.25}
\psfig{file=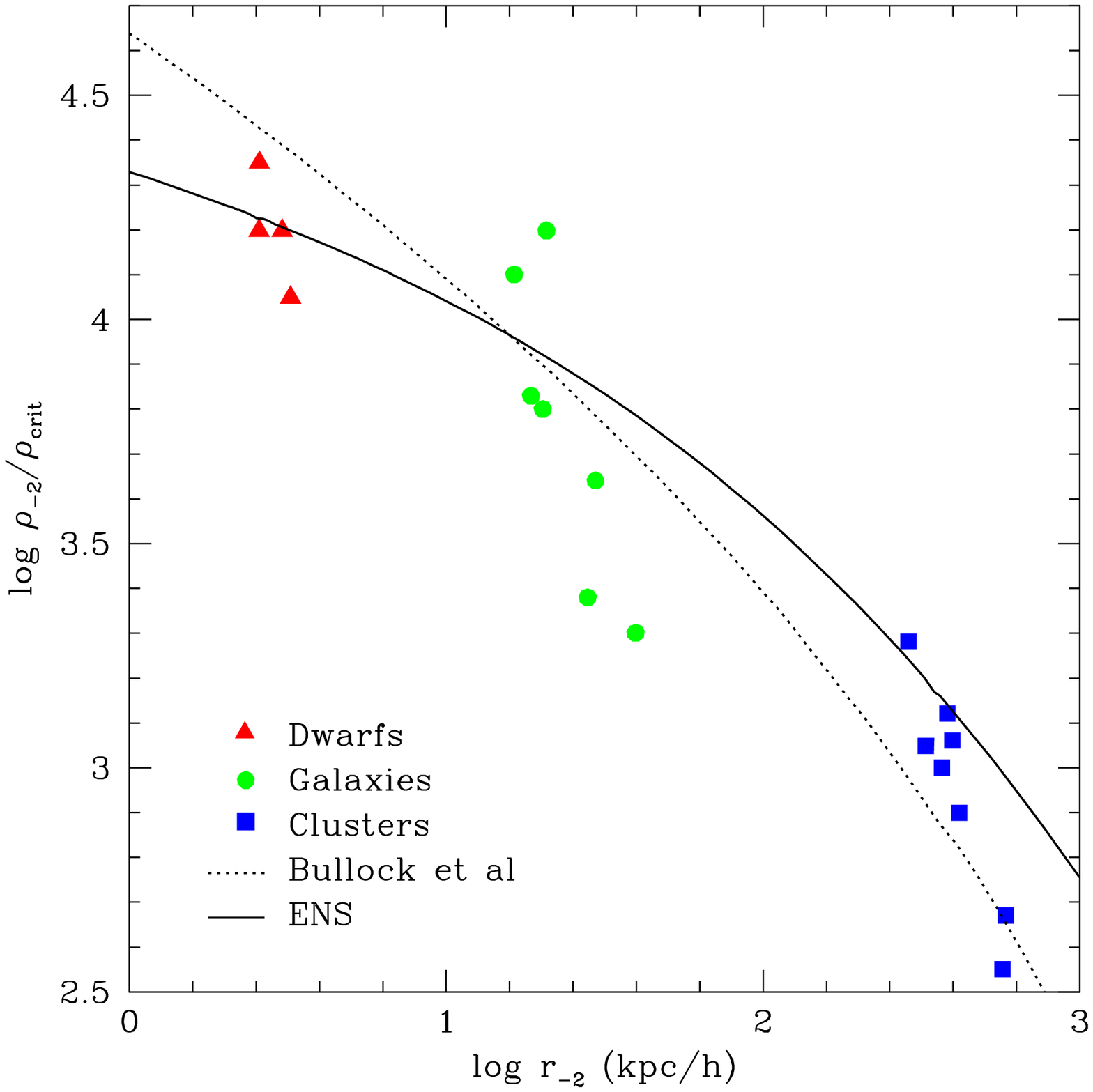, width=180mm}
\end{center}
\caption{
The radius, $r_{-2}$, where the logarithmic slope of the density profile takes
the ``isothermal'' value, $\beta(r_{-2})=2$, plotted versus the local density at
that radius, $\rho_{-2}=\rho(r_{-2})$, for all simulated halos in our
series. This figure illustrates the mass dependence of the central concentration
of dark matter halos: low mass halos are systematically denser than their more
massive counterparts. Solid and dotted lines indicate the scale
radius-characteristic density correlation predicted by the formalisms presented
by Eke, Navarro \& Steinmetz (2001) and Bullock et al (2001). These parameters
may be used, in conjunction with eq.~\ref{eq:newprof}, to predict the mass
profile of $\Lambda$CDM halos.
\label{figs:rm2rhom2}}
\end{figure*}

We have used the ENS and B01 formalisms to predict the halo mass dependence of
the scaling parameters, $\rho_{-2}$ and $r_{-2}$, and we compare the results
with our simulations in Figure~\ref{figs:rm2rhom2}. The ENS prediction is shown
by the solid line whereas the dotted line shows that of B01. Both formalisms
reproduce reasonably well the trend seen in the simulations, so that one can use
either, in conjunction with eq.~\ref{eq:newprof} (with $\alpha$ in the range
0.1-0.2), to predict the structure of a $\Lambda$CDM halo. A simple code that
computes $r_{-2}$ and $\rho_{-2}$ as a function of mass in various cosmological
models is available upon request from the authors. Existing codes that compute
NFW halo parameters as a function of mass and of other cosmological parameters
may also be used, noting that $\rho_{-2}=\rho_s/4$ and that $r_{-2}=r_s$.

Finally, we note that neither formalism captures perfectly the mass dependence
of the characteristic density; small but significant deviations, as well as a
sizable scatter, are evident in Figure~\ref{figs:rm2rhom2}. Dwarf galaxy halos
appear to be less concentrated than predicted by the formalism proposed by B01;
a similar observation applies to cluster halos when compared to ENS'
predictions. Such shortcomings should be considered when deriving cosmological
constraints from fits to observational data (see, e.g., Zentner \& Bullock 2002,
McGaugh et al 2003); and suggest again that direct comparison between
observation and simulations is preferable to the use of fitting formulae.

\section{Summary}
\label{sec:conc}

We have analyzed the mass profile of $\Lambda$CDM halos in a series of
simulations of high mass, spatial, and temporal resolution. Our series targets
halos spanning five decades in mass: ``dwarf'' galaxy halos with virial circular
velocities of order $V_{200}\sim 30$ km s$^{-1}$; ``galaxy''-sized halos with
$V_{200} \sim 200$ km s$^{-1}$; and ``cluster'' halos with $V_{200} \sim 1200$
km s$^{-1}$.  Each of the nineteen halos in our series was simulated with
comparable numerical resolution: they have between $8\times 10^5$ and $4\times
10^6$ million particles within the virial radius, and have been simulated
following the ``optimal'' prescription for time-stepping and gravitational
softening laid down in the numerical convergence study of P03.

The high resolution of our simulations allows us to probe the inner properties
of the mass profiles of $\Lambda$CDM halos, down to $\sim 0.5\%$ of $r_{200}$ in
our best resolved runs. These results have important implications for the
structure of the inner cusp in the density profile and resolve some of the
disagreements arising from earlier simulation work. Our main conclusions may be
summarized as follows.

\begin{itemize}

\item $\Lambda$CDM halo density profiles are ``universal'' in shape:
i.e., a simple fitting formula reproduces the structure of all
simulated halos, regardless of mass. Both the NFW
profile and the profile proposed by M99 describe the density and
circular velocity profiles of simulated halos reasonably well. Best
NFW fits to the circular velocity profiles deviate by less than $10\%$
over the region which is well resolved numerically. Best M99 fits
reproduce circular velocity profiles to better than $25\%$ over the
same region. It should be noted, however, that the deviations increase
inwards and are typically maximal at the innermost resolved radius, a
result that warns against extrapolating to smaller radii with these 
fitting formulae.

\item $\Lambda$CDM halos appear to be ``cuspy'': i.e., the dark matter
density increases monotonically towards the center with no evidence for
a well-defined ``core'' of constant density. We find no evidence, however,
for a central asymptotic power-law bin the density profiles. These
become progressively shallow inwards and are significantly shallower
than isothermal at the innermost resolved radius, $r_{\rm conv}$. At
$r\sim 0.01 \, r_{200}$, the average slope of ``cluster'',
``galaxy'' and ``dwarf'' halos halos is $\beta(r_{\rm conv}) \sim
1.1$, $\sim 1.2$, and $\sim 1.35$, respectively. This is steeper than
predicted by the NFW profile but shallower than the asymptotic slope
of the M99 profile.

\item 
The density and enclosed mass at $r_{\rm conv}$ may be used to derive an upper
limit on any asymptotic value of the inner slope. Cusps as steep as
$\beta_0=1.5$ are confidently ruled out in essentially all cases; the asymptotic
slope of the NFW profile ($\beta_0=1$) is still consistent with our data.  The
radial dependence of $\beta(r)$ differs from that of the NFW profile, however,
decreasing more slowly with decreasing radius than is predicted. For some
scalings of the NFW fitting formula to the numerical data, this shape difference
appears as a dark matter ``excess'' near the center which has (erroneously) been
interpreted indicating a steeply divergent density cusp.

\item 
A simple formula where $\beta(r)$ is a power law of radius reproduces
the gradual radial variation of the logarithmic slope and its apparent
failure to converge to any specific asymptotic value
(eq.~\ref{eq:newprof}). This formula leads to much improved fits to
the density profiles of simulated halos, and may prove a safer choice
when comparison with observation demands extrapolation below the
innermost converged radii of the simulations.

\end{itemize}

Our study demonstrates that, although simple fitting formulae such as NFW are
quite accurate in describing the global structure of $\Lambda$CDM halos, one
should be aware of the limitations of these formulae when interpreting
observational constraints. Extrapolation beyond the radial range where these
formulae have been validated is likely to produce substantial errors. Proper
account of the substantial scatter in halo properties at a given halo mass also
appears necessary when assessing the consistency of observations with a
particular cosmological model. Direct comparison between observations and
simulations (rather than with fitting formulae) is clearly preferable whenever
possible. Given the computational challenge involved in providing consistent,
robust, and reproducible theoretical predictions for the inner structure of CDM
halos it is likely that observational constraints will exercise to the limit our
hardware and software capabilities for some time to come.


This work has been supported by computing time generously provided by the High
Performance Computing Facility at the University of Victoria, as well as by the
Edinburgh Parallel Computing Centre and by the Institute for Computational
Cosmology at the University of Durham. Expert assistance by Colin Leavett-Brown
in Victoria and Lydia Heck in Durham is gratefully acknowledged. JFN is
supported by the Alexander von Humboldt Foundation, the Natural Sciences and
Engineering Research Council of Canada, and the Canadian Foundation for
Innovation.

\begin{table*}
\begin{center}
\caption{Parameters of the parent cosmological simulations}
\begin{tabular}{llclcc}
Label & $L_{\rm box}$ & $z_i$ & $m_p$ & $\epsilon$ & CODE \\ & [$h^{-1}$ Mpc] &
& [$h^{-1} \, M_{\odot}$] & [$h^{-1}$ kpc] & \\
\hline
ENS01           & $32.5$   & $49.0$ & $1.36\times 10^9$    & $10$ & {\tt AP3M} \\
SGIF-128           & $35.325$ & $49.0$ & $1.75\times 10^9$    & $10$ & {\tt GADGET} \\
$\Lambda$CDM-512  & $479.0$    & $36.0$ & $6.82\times 10^{10}$ & $30$ & {\tt GADGET} \\
\label{tab:numexp}
\end{tabular}
\end{center}
\end{table*}

\begin{table*}
\begin{center}
\caption{Main parameters of resimulated halos}
\begin{tabular}{cclrrrrcc}
Label & $z_i$ & $\epsilon$ &$N_{200}$ & $M_{200}$ & $r_{200}$ & $V_{200}$ &
$r_{\rm conv}$ & CODE \\
      &     & [$h^{-1}$ kpc] &      &  [$h^{-1} \, M_{\odot}$] & [$h^{-1}$ kpc] &  [km
s$^{-1}$]  & [$h^{-1}$ kpc] &  \\
\hline
D1 & $74$ & $0.0625$  & $784980$  & $7.81\times 10^{9}$  & $32.3$   & $32.3$   & $0.34$& {\tt GADGET}\\
D2 & $49$ & $0.0625$  & $778097$  & $9.21\times 10^{9}$  & $34.1$   & $34.1$   & $0.37$& {\tt GADGET}\\
D3 & $49$ & $0.0625$  & $946421$  & $7.86\times 10^{9}$  & $32.3$   & $32.3$   & $0.33$& {\tt GADGET}\\
\smallskip
D4 & $49$ & $0.0625$  & $1002098$ & $9.72\times 10^{9}$  & $34.7$   & $34.7$   & $0.32$& {\tt GADGET}\\
G1 & $49$ & $0.15625$ & $3447447$ & $2.29\times 10^{12}$ & $214.4$  & $214.4$  & $1.42$& {\tt GADGET}\\
G2 & $49$ & $0.5$     & $4523986$ & $2.93\times 10^{12}$ & $232.6$  & $232.6$  & $1.25$& {\tt PKDGRAV}\\
G3 & $49$ & $0.45$    & $2661091$ & $2.24\times 10^{12}$ & $212.7$  & $212.7$  & $1.65$& {\tt PKDGRAV}\\
G4 & $49$ & $0.3$     & $3456221$ & $1.03\times 10^{12}$ & $164.0$  & $164.0$  & $1.01$& {\tt PKDGRAV}\\
G5 & $49$ & $0.35$    & $3913956$ & $1.05\times 10^{12}$ & $165.0$  & $165.0$  & $1.02$& {\tt PKDGRAV}\\
G6 & $49$ & $0.35$    & $3739913$ & $9.99\times 10^{11}$ & $162.5$  & $162.5$  & $1.03$& {\tt PKDGRAV}\\
\smallskip
G7 & $49$ & $0.35$    & $3585676$ & $9.58\times 10^{11}$ & $160.3$  & $160.3$  & $1.02$& {\tt PKDGRAV}\\
C1 & $36$ & $5.0$     & $1565576$ & $7.88\times 10^{14}$ & $1502.1$ & $1502.1$ & $16.8$& {\tt GADGET}\\
C2 & $36$ & $5.0$     & $1461017$ & $7.36\times 10^{14}$ & $1468.1$ & $1468.1$ & $16.9$& {\tt GADGET}\\
C3 & $36$ & $5.0$     & $1011918$ & $5.12\times 10^{14}$ & $1300.6$ & $1300.6$ & $16.1$& {\tt GADGET}\\
C4 & $36$ & $5.0$     & $1050402$ & $5.31\times 10^{14}$ & $1316.7$ & $1316.7$ & $15.9$& {\tt GADGET}\\
C5 & $36$ & $5.0$     & $1199299$ & $6.05\times 10^{14}$ & $1375.5$ & $1375.5$ & $16.2$& {\tt GADGET}\\
C6 & $36$ & $5.0$     & $1626161$ & $8.19\times 10^{14}$ & $1521.1$ & $1521.1$ & $15.5$& {\tt GADGET}\\
C7 & $36$ & $5.0$     & $887837$  & $4.50\times 10^{14}$ & $1245.8$ & $1245.8$ & $16.4$& {\tt GADGET}\\
C8 & $36$ & $5.0$     & $1172850$ & $5.92\times 10^{14}$ & $1365.4$ & $1365.4$ & $16.8$& {\tt GADGET}\\
\label{tab:halpar}
\end{tabular}
\end{center}
\end{table*}

\begin{table*}
\begin{center}
\caption{Fit and structural parameters of resimulated halos}
\begin{tabular}{cccccccccc}
Label & $r_{-2}$ & $\rho_{-2}$ & $r_{\rm max}$ & $V_{\rm max}$ & $r_{s}$ & $\rho_s$ & $r_M$ & $\rho_M$ & $\alpha$\\
      & [$h^{-1}$ kpc] &  [$\rho_{\rm crit}$] & [$h^{-1}$ kpc] &  [km s$^{-1}$]
& [$h^{-1}$ kpc] & [$\rho_{\rm crit}$] &  [$h^{-1}$ kpc] & [$\rho_{\rm crit}$] &\\
\hline
D1 & $3.23$   &$1.12e4$ &$6.07$  &$39.1$ &$2.59$ &$7.03e4$&$5.38$ &$7.58e3$ & $0.164$\\
D2 & $3.04$   &$1.58e4$ &$6.64$  &$44.2$ &$2.43$ &$9.61e4$&$2.27$ &$1.17e5$ & $0.211$\\
D3 & $2.57$   &$1.58e4$ &$6.35$  &$36.9$ &$2.94$ &$5.01e4$&$4.05$ &$2.51e4$ & $0.122$\\
\smallskip
D4 & $2.57$   &$2.24e4$ &$4.27$  &$45.7$ &$2.06$ &$1.49e5$&$2.18$ &$1.36e5$ & $0.166$\\
G1 & $18.5$   &$6.76e3$ &$23.7$  &$1.95e2$ &$23.2$ &$4.06e4$&$19.4$ &$8.43e4$ & $0.142$\\
G2 & $28.0$   &$2.40e3$ &$68.5$  &$1.78e2$ &$16.8$ &$1.13e5$&$19.4$ &$8.43e4$ & $0.191$\\
G3 & $20.2$   &$6.31e3$ &$43.4$  &$1.96e2$ &$28.0$ &$1.52e4$&$47.3$ &$8.24e3$ & $0.142$\\
G4 & $29.6$   &$4.37e3$ &$63.4$  &$2.49e2$ &$12.3$ &$6.78e4$&$16.8$ &$3.44e4$ & $0.177$\\
G5 & $20.7$   &$1.58e4$ &$67.7$  &$2.91e2$ &$13.8$ &$5.20e4$&$15.3$ &$4.23e4$ & $0.184$\\
G6 & $39.6$   &$2.00e3$ &$96.4$ &$2.26e2$ &$15.3$ &$3.79e4$&$20.7$ &$2.03e4$ & $0.171$\\
\smallskip
G7 & $16.4$   &$1.26e4$ &$29.9$  &$1.94e2$ &$13.4$ &$6.22e4$&$14.9$ &$5.15e4$ & $0.138$\\
C1 & $5.84e2$ &$4.68e2$ &$1.03e3$&$1.48e3$ &$440$  &$3.36e3$&$661$  &$1.58e3$ & $0.133$\\
C2 & $3.95e2$ &$1.15e3$ &$9.99e2$&$1.51e3$ &$362$  &$5.17e3$&$396$  &$4.46e3$ & $0.215$\\
C3 & $3.27e2$ &$1.12e3$ &$6.15e2$&$1.38e3$ &$249$  &$9.07e3$&$278$  &$7.44e3$ & $0.188$\\
C4 & $4.16e2$ &$7.94e2$ &$6.57e2$&$1.38e3$ &$315$  &$5.47e3$&$339$  &$4.91e3$ & $0.161$\\
C5 & $2.87e2$ &$1.91e3$ &$6.48e2$&$1.42e3$ &$271$  &$8.45e3$&$326$  &$5.88e3$ & $0.215$\\
C6 & $3.82e2$ &$1.32e3$ &$6.94e2$&$1.64e3$ &$297$  &$8.70e3$&$302$  &$8.75e3$ & $0.203$\\
C7 & $5.69e2$ &$3.55e2$ &$1.25e3$&$1.25e3$ &$283$  &$3.92e3$&$475$  &$2.11e3$ & $0.129$\\
C8 & $3.68e2$ &$1.00e3$ &$9.35e2$&$1.44e3$ &$361$  &$4.41e3$&$345$  &$5.04e3$ & $0.219$\\
\label{tab:fitpar}
\end{tabular}
\end{center}
\end{table*}

\newpage

\epsscale{1.0}
\epsscale{1.0}

\epsscale{1.0}

\epsscale{1.0}

\epsscale{1.0}

\epsscale{1.0}
\epsscale{1.0}

\epsscale{1.0}

\end{document}